\documentclass{article}
\usepackage[letterpaper,top=2cm,bottom=2cm,left=3cm,right=3cm,marginparwidth=1.75cm]{geometry}
\usepackage{graphicx}
\usepackage{amsmath}
\usepackage{graphicx}
\PassOptionsToPackage{hyphens}{url}
\usepackage[colorlinks=true, allcolors=blue]{hyperref}
\usepackage{makecell}
\usepackage{booktabs}
\usepackage{tabularx}
\usepackage{ltablex}
\usepackage{caption}
\usepackage{subcaption}
\usepackage{stackengine}
\usepackage{multirow}
\usepackage{authblk}
\usepackage[table]{xcolor}
\usepackage{abstract}
\usepackage{breakcites}
\usepackage{lineno}

\title{Who is driving the conversation? Analysing the nodality of British MPs and journalists on social media}
\author[1,2,*,$\dagger$]{Sukankana Chakraborty}
\author[1,*]{Leonardo Castro-Gonzalez}
\author[1,3]{Helen Margetts}
\author[1,4]{Hardik Rajpal}
\author[1]{Daniele Guariso}
\author[1]{Jonathan Bright}

\affil[1]{Public Policy Programme, The Alan Turing Institute, British Library, 96 Euston Rd, London, NW1 2DB, UK.}
\affil[2]{Centre for Advanced Spatial Analysis, University College London, Gower Street, London, WC1E 6BT, UK}
\affil[3]{Oxford Internet Institute, University of Oxford, 1 St Giles, Oxford, OX1 3JS, UK}
\affil[4]{Department of Mathematics, Imperial College London, South Kensington Campus, London, SW7 2AZ, UK}
\affil[*]{These authors contributed equally}
\affil[$\dagger$]{corresponding author: \url{sukankana.chakraborty@ucl.ac.uk}}

\date{}
\begin{document}

\maketitle


\begin{abstract}
With the rise of social media, political conversations now take place in more diffuse environments. In this context, it is not always clear why some actors, more than others, have greater influence on how discussions are shaped. To investigate the factors behind such influence, we build on ``nodality” a concept in political science which describes the \textit{capacity} of an actor to exchange information within discourse networks. This concept goes beyond traditional network metrics that describe the position of an actor in the network to include exogenous drivers of influence (e.g. factors relating to organisational hierarchies). 
We study online discourse on Twitter (now X) in the UK to measure the relative nodality of two sets of policy actors - Members of Parliament (MPs) and accredited journalists - on four policy topics. We find that influence on the platform is driven by two key factors: (i) ``active nodality", derived from the actor’s level of topic-related engagement and (ii) ``inherent nodality" which is independent of the platform discourse and reflects the actor's institutional position. These findings significantly further our understanding of the origins of influence on social media platforms, and suggest in which contexts influence is transferable across topics. 
\end{abstract}

\maketitle
\section{Introduction}
\label{sec:introduction}
In recent decades, internet-based platforms have transformed policy discourse significantly by reducing the costs of exchanging information and by providing a more democratised environment for discussions in which individuals from diverse backgrounds - elite politicians and members of the public alike - can engage in conversations. This has led to interesting, complex online dynamics in which some actors, often driven by factors that are poorly understood, wield greater influence than others on the course of the conversation \cite{harris2024perceived,bright_social_2016, bright_does_2020, sprejer_actor-based_2022, parmelee_political_2013, usherwood_sticks_2017}. 
Understanding these drivers of influence is crucial in order to better gauge which actors have the capacity to shape policy debates and online narratives \textemdash in some cases spontaneously leading to significant and undesirable real-world outcomes (such as the UK riots of 2024 \cite{boukari2024far}).
Although, much research has been dedicated to understanding the relative influence of actors on social media \cite{dubois_multiple_2014, beguerisse2014interest, anger_measuring_2011, bakshy_everyones_2011, mei_finding_2015, riquelme_measuring_2016, esteve_del_valle_leaders_2018, rezaie_measuring_2020, kolic_quantifying_2022, flamino_political_2023, zhang_exploratory_2023} \textemdash much of this work is centred around communications literature which uses traditional network metrics to examine the roles of actors in the online network, rarely examining external factors - such as institutional or individual characteristics - that could potentially drive differential online attention towards these actors. Particularly in the context of political actors, a long-standing question has been \textit{to what extent does the eliteness of an actor contribute to their role in online discussions?} \cite{meraz2009there,steinberg2016central,barbera2019leads}. 
For instance, do frontbench Members of Parliament (MPs) always attract greater online attention than backbench MPs, therefore giving them a bigger say in the policy discourse? Or can individual factors, such as active engagement with a specific issue within their constituency outweigh the significance of institutional status?

With this is in mind, here we tackle the question of ‘who drives political discourse, and how?’ through the lens of nodality \cite{castelnovo_nodality_2021, howlett_government_2009, howlett_information_2022}. Nodality is a concept in political science that goes beyond simple network metrics and describes the \textit{capacity} of an actor to be at the centre of information networks \cite{hood_tools_1983,Hood2008,margetts_nodality_2023}.
Nodality used to be something that government \textemdash as a set of organisations \textemdash possessed almost by virtue of being the ``water mill'' at the centre of society’s information networks \cite{margetts_nodality_2023}. 
But in a digital era, where policy debates take place in more open and diffuse information environments, governmental actors face greater competition for nodality \cite{castelnovo_nodality_2021}. Widespread use of the internet means that non-governmental organisations or individuals such as journalists, politicians, public figures and even citizens themselves can acquire nodality, which can challenge the government’s capacity to capture public attention and drive policy and media agendas \cite{margetts_political_2019, howlett_government_2009, howlett_information_2022,shearing_nodal_2003}. 
So far there has been little effort to quantify nodality, or the extent to which an actor can influence a discourse, beyond simple metrics derived from network structure alone \cite{margetts_governing_2006, petricek_web_2006}.

Framed within the context of policy-related discourse in the UK, here we propose a method that measures the relative nodality of policy-actors as a means to determine the extent to which they shape policy conversations. We construct a discourse network for each of the topics and a corresponding null network which captures all the interactions that occur outside the topic. We measure the relative nodality of actors, specifically their inherent and active nodalities by comparing their centrality metrics across these two networks in a principal component analysis (PCA). We assume that inherent nodality is high when centrality metrics are high across both networks, and active nodality is high when the centrality of the actor is high in the topic network but low in the null network. In an independent group-level transfer entropy (TE) analysis we show how the mean nodalities of a group of actors relate to their capacity to drive activity .  
This method is then used to analyse a subset of political actors in the UK: Members of Parliament (MPs) and accredited journalists, as both MPs and journalists play key roles in the British policy system. More specifically, we examine to what extent institutional and individual characteristics drive their interactions with each other. 
We study conversations around four policy topics in the year 2022: (1) the Russia-Ukraine war (2) the challenging economic situation facing ordinary citizens in the UK (Cost-of-Living crisis) (3) Britain’s departure from the EU (Brexit) and (4) the aftermath of the COVID-19 pandemic. All four issues were contested in policy terms and related to an increasingly challenging economic situation in 2022, driven by Brexit \cite{diamond_post-brexit_2023, dhingra_expecting_2022}, the pandemic \cite{blundell_inequality_2022}, the Russia-Ukraine war \cite{mbah_russian-ukraine_2022, liadze_economic_2023} and the resulting energy crisis \cite{commons_crisis_2022_1, commons_crisis_2022_2}, as well as a ``mini-budget'' in the autumn of 2022 which led to sharp and sustained rises in interest rates \cite{chadha_monetary_2022, tosun_growth_2023, li_britains_2023}. 

For the purpose of our study, we examine the nodality of actors along two key dimensions: \textbf{(i)} inherent nodality, which is earned by virtue of one’s institutional role (such as a frontbench position) and \textbf{(ii)} active or ``topic specific'' nodality, which is earned through engagement on a particular topic. We show that inherent nodality \textemdash which is independent of the platform \textemdash is a useful indicator of the actor's influence on the platform, but that both dimensions of nodality (inherent and active) significantly impact the capacity of an actor to influence activity (e.g., drive activity of others) on the platform. 
Analysing these two types of nodality is crucial to understanding the dynamics of influence in real-world policy networks. By distinguishing the influence that is derived from institutional status, we are able to determine the \emph{nodal advantage} of actors across topics, but also uncover early signs of emerging policy actors in specific topics, such as the opposition backbencher Zarah Sultana, who figured prominently in our early analysis of the cost-of-living crisis discourse, and during the 2024 general election was recognised as the most influential MP on TikTok\footnote{\href{https://www.standard.co.uk/news/politics/zarah-sultana-labour-mp-coventry-south-tiktok-b1159562.html}{standard.co.uk/news/politics/zarah-sultana-labour-mp-coventry-south-tiktok-b1159562.html}}.


%

\section{Data}
\label{ssec:data}
The data for our study was collected using the Twitter API. We study 6M tweets that were published by UK MPs and journalists between January 14, 2022 and January 13, 2023. 
Of the 650 House of Commons MPs at the time, we found that 581 MPs have active Twitter accounts and a total of 580,312 tweets were collected from these accounts. We also obtained a list of UK journalists and their Twitter handles from \hyperlink{https://journalism.co.uk/prof}{journalism.co.uk}. Of the 9594 journalists on the list, we were able to collect tweets from 8606 journalist accounts and over 5M tweets were collected from these accounts.

Our study examines the discourse between MPs and journalists on four policy topics: Russia-Ukraine War, cost-of-living crisis, Brexit, COVID-19. Tweets were classified into each of these topics using a weak supervision approach \cite{Ratner2019, castro-gonzalez_cheap_2024}. Details of the approach can be found in the \ref{sec:appendix}. Tweets that were not assigned to any of the above topics were left unlabelled. 
We visualise the discourse on each topic by building a weighted, directed graph where nodes represent actors and any interaction from an actor \emph{j} to an actor \emph{i}, such as a retweet, mention or reply, is captured as an edge $i\to j$. As nodality describes the capacity of a node to transmit information, we consider any interaction in the form of retweets, mentions and replies from actor $j$ to $i$ as a flow of information from $i \to j$.
The edge weight $w_{ij}$ corresponds to the number of times a target node interacted with the source node, thus capturing the volume of information passing from the source actor to the target actor.

\section{Results}
\label{sec:quantify_nodality}
\subsection{The multi-dimensional nature of influence}
\label{ssec:zarah_sultana}
We begin by qualitatively analysing the discourse networks for each topic. To do this, we employ centrality metrics \textemdash a key concept in network analysis which describes the position of an actor in a social network  \cite{anger_measuring_2011, bakshy_everyones_2011, mei_finding_2015, riquelme_measuring_2016, rezaie_measuring_2020, zhang_exploratory_2023, essaidi_new_2020}. 
In particular, we examine the outdegree of the actors in the discourse network as an indication of how much information they are able to transmit in the rest of the network. 

\begin{figure*}[ht]
     \centering
     \includegraphics[width=\textwidth]{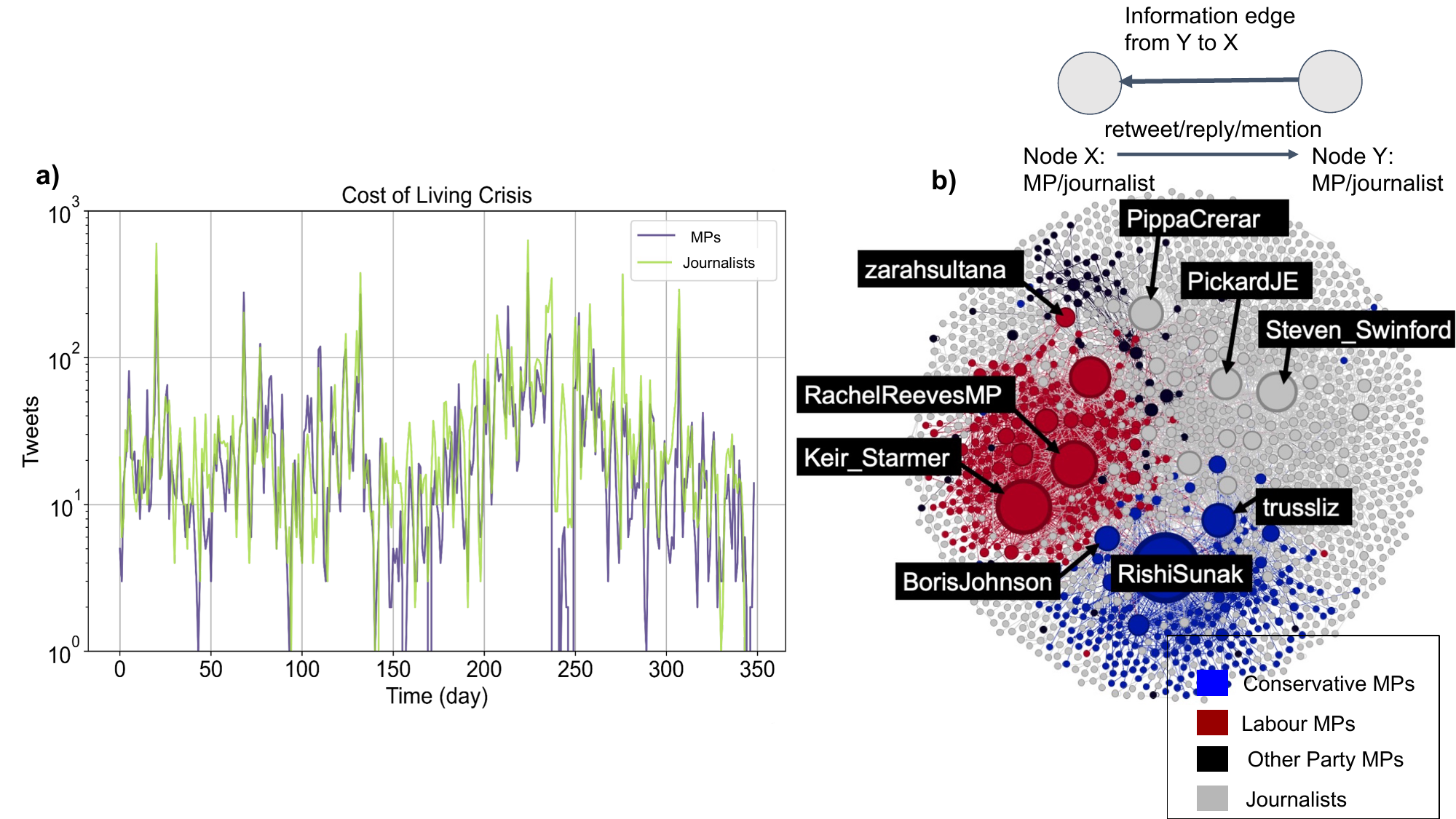}
     \caption{\label{fig:CoL_nodality}Qualitative analysis of the cost-of-living crisis discourse. Figure (a) shows the daily tweet volumes from MP and Journalist accounts related to the cost-of-living crisis discussion between January 14,2022 and January 13, 2023. We generally observe similar levels of activity from both actors. 
     Figure (b) shows the giant connected component of the cost-of-living crisis conversation network. Nodes representing journalists are shown in grey, while MPs are represented using coloured nodes where the colour signifies the political party they are affiliated with (Conservative, Labour, other parties). Edges here take the colour of the target node. The force-layout algorithm was used to visualise the network and we find that nodes appear clustered based on party affiliations. The size of a node corresponds to the weighted out-degree which reflects the number of times a user has been retweeted (replied to or mentioned) by others in the network. Some actors who stand out in the network and appear to be central to high levels of activity on the topic have been explicitly labelled.}
\end{figure*}

Our analysis shows that some nodes \textemdash typically MPs with front-bench positions \textemdash appear prominent in all four conversations. 
As an example we present the cost-of-living crisis network in Figure \ref{fig:CoL_nodality}. We find that nodes with the highest outdegrees, i.e., with the most engagement, are MPs with important parliamentary positions i.e. Cabinet and Shadow Cabinet MPs during 2022-2023, such as Boris Johnson, Rishi Sunak, Liz Truss, Keir Starmer, Rachel Reeves and Ed Milliband, and political editors at leading UK newspapers, such as Pippa Crerar (from The Guardian), Steven Swindon (The Times), and Jim Pickard (Financial Times). 
Rishi Sunak has the highest outdegree in the network which can be attributed to the two key roles he held during the year 2022-23 \textemdash the Chancellor of the Exchequer, a pivotal role in public spending, and eventually that of the Prime Minister of the UK. We find that Rachel Reeves, the Shadow Chancellor of the Exchequer, and Nadhim Zahawi, who was the Chancellor under Liz Truss’ premiership in 2022, are also among the key actors in the cost-of-living crisis discussion. 

At the same time, we find that some actors appear prominent only in specific discourse networks.
For instance, in the case of the cost-of-living crisis discussion we find that Zarah Sultana, a Backbench Labour MP, despite their less significant role in the parliament plays a key role in the conversation (Figure \ref{fig:CoL_nodality}).
Furthermore, we find that their centrality in the cost-of-living crisis discussion does not extend to any of the other topics studied here. Whereas actors with more senior positions within the Cabinet (or shadow frontbench) in 2022 such as Boris Johnson, Rishi Sunak, Liz Truss and Keir Starmer appear consistently central in all four topics of discussion, suggesting that influence on these platforms can be multi-dimensional \cite{dubois_multiple_2014}.

While centrality metrics are useful for qualitatively analysing an actor's influence in online discourse, it ultimately falls short in determining factors driving such influence, such as to what extent does an actor's institutional status or level of engagement on a topic contribute to their influence on the platform?
We rely on the concept of nodality to answer these questions \cite{margetts_nodality_2023}. Nodality describes the capacity of an actor to be in the centre of an information network and thus draws out factors driving an actor's influence on a platform, specifically by capturing multiple dimensions of influence such as extroversion (from engagement on the platform), visibility (from institutional status) etc., which cannot be fully described by any one centrality measure but can be captured by using a combination of centrality metrics. Authors in prior work show how each centrality metric relates to different dimensions of nodality \cite{margetts_governing_2006, petricek_web_2006}, and building on this work, we establish a methodology that combines several centrality measures to quantify the relative nodality of actors in a discourse network. We measure nodality along two dimensions \textemdash inherent nodality which tells us the extent to which an actor's institutional role contributes to their influence on the platform and active nodality which highlights the topic-related influence the actor has on the platform.

We consider six centrality metrics in our work \textemdash degree centrality, betweenness centrality, Eigenvalue centrality, Authority and Hub score \cite{kleinberg_authoritative_1999} and finally the strength of a node. Each of these metrics describe dimensions of nodality such as extroversion, authority and visibility of actors, which are key to determining the capacity of actors to transmit information \cite{margetts_governing_2006}.
In addition to the traditional centrality measures, we add two new metrics that are specific to the Twitter platform and that explore the follower count of users to estimate their popularity and reach on the platform \cite{riquelme_measuring_2016}. 
We term these metrics: (i) the funnel bandwidth and (ii) the amplification bandwidth of an actor. The funnel bandwidth of an actor is a measure of the extent to which they facilitate the passage of information in the network. This is quantified in terms of the volume of interactions (retweets, mentions, replies), weighted by the actor's follower count. Amplification bandwidth, on the other hand, measures the exposure of information from a user in two hops. We measure this as the number of times a user $i$ interacts with others, weighted by the number of followers of each target user $j$. More specifically, these metrics are defined respectively as:
\begin{align}
    \label{eq:new_measures}
    \nu^{i} = \frac{1}{\langle k_\mathrm{in}\rangle}\sum_hf^iw_{hi} \hspace{10pt} \mathrm{and} \hspace{10pt}
    \mu^{i} = \frac{1}{\langle k_\mathrm{out}\rangle} \sum_j f^jw_{ij}.
\end{align}
Here $f^i$ represents the number of followers of node $i$, $w_{ij}$ is the weight of link $i\to j$, $\langle k_\mathrm{in}\rangle$ is the average in-degree of the weighted network and $\langle k_\mathrm{out}\rangle$ is the average out-degree.

\subsection{Inherent and active nodality of individual actors}
\label{ssec:methodology}
Our aim is to measure the relative nodality of actors along two dimensions: institutional nodality and topic-related (active) nodality. 
To do this, we first construct two networks for each topic \textemdash the topic-related discourse network and a corresponding null network that captures all the interactions outside the topic. In order to capture the inherent and active nodality of actors we compare their centrality measures across the topic and the null network. More specifically, since inherent nodality is driven by institutional status and thus is independent of the topic, we assume that inherent nodality is high for an actor when centrality measures are high across both topic and null networks. Similarly, we assume that active (or topic-specific) nodality is high when centrality measure is low in the null network but high in the topic network, i.e. the actor has no `eliteness' on the platform but has gained nodality only on the specific topic. 

To quantify the two types of nodality, we perform a principal component analysis (PCA) on an $n$ x $2m$ matrix with rows corresponding to $n$ nodes, and columns corresponding to $m$ centrality metrics (computed for each network and so $2m$ columns in total). We enforce conditions for inherent and active nodality by introducing an eigenvector test. The eigenvector test checks the first two eigenvectors of the PCA transformation to verify that in the case of inherent nodality, centrality measures across both networks are all positively (or all negatively) correlated with the first principal component. In the case of active nodality we check if the centrality measures on the topic network are positively correlated with the second principal component and centrality measures on the null network are negatively correlated to the second principal component. 
If for example we perform a PCA with any three centrality measures, such that the eigenvectors in the PCA transformation is given by $\boldsymbol{e}=(f_1^t, f_2^t,f_3^t, f_1^0, f_2^0, f_3^0)$, where $f_n$ refers to the centrality measures and the superscript $t/0$ indicates the type of network, i.e. topic or null network. Then we say a configuration passes the eigenvector test if the first eigenvector is of the form $\boldsymbol{e_1}=(1, 1, 1, 1, 1, 1)$ or $\boldsymbol{e_1}=(-1, -1, -1, -1, -1, -1)$ and the second eigenvector is of the form $\boldsymbol{e_2}=(-1, -1, -1, 1, 1, 1)$ or $\boldsymbol{e_t}=(1, 1, 1, -1, -1, -1)$. At this stage, we generate all combinations of three or more centrality metrics and only retain combinations that pass the eigenvector test (i.e. satisfy our conditions of inherent and active nodality).  

Next, to determine the set of centrality measures that best captures inherent and active nodality of actors we apply an additional condition to the PCA-transformed data. Given that inherent nodality is topic-agnostic, we assume that the group of actors with high relative inherent nodality will be similar across all four topics. 
To apply this condition, we use a $k-$means clustering method on the PCA-transformed data to cluster nodes based on their relative nodalities. We find that $k=3$ optimally divides the population along the first principal component axis, i.e. based on their inherent nodality. 
More specifically, we obtain three clusters: actors with inherent nodality (i) higher than the mean of the population, (ii) same as the mean and (iii) lower than the mean.
We compare the cluster of actors with high inherent nodality across all four topics. 
The combination of metrics that maximises the number of shared actors across all four topics includes: the strength of a node, their degree centrality and the funnel bandwidth of the node. In the rest of the paper, we use this set of centrality metrics to measure the relative inherent and active nodalities of actors.

\begin{figure*}[ht]
     \centering
     \includegraphics[width=\textwidth]{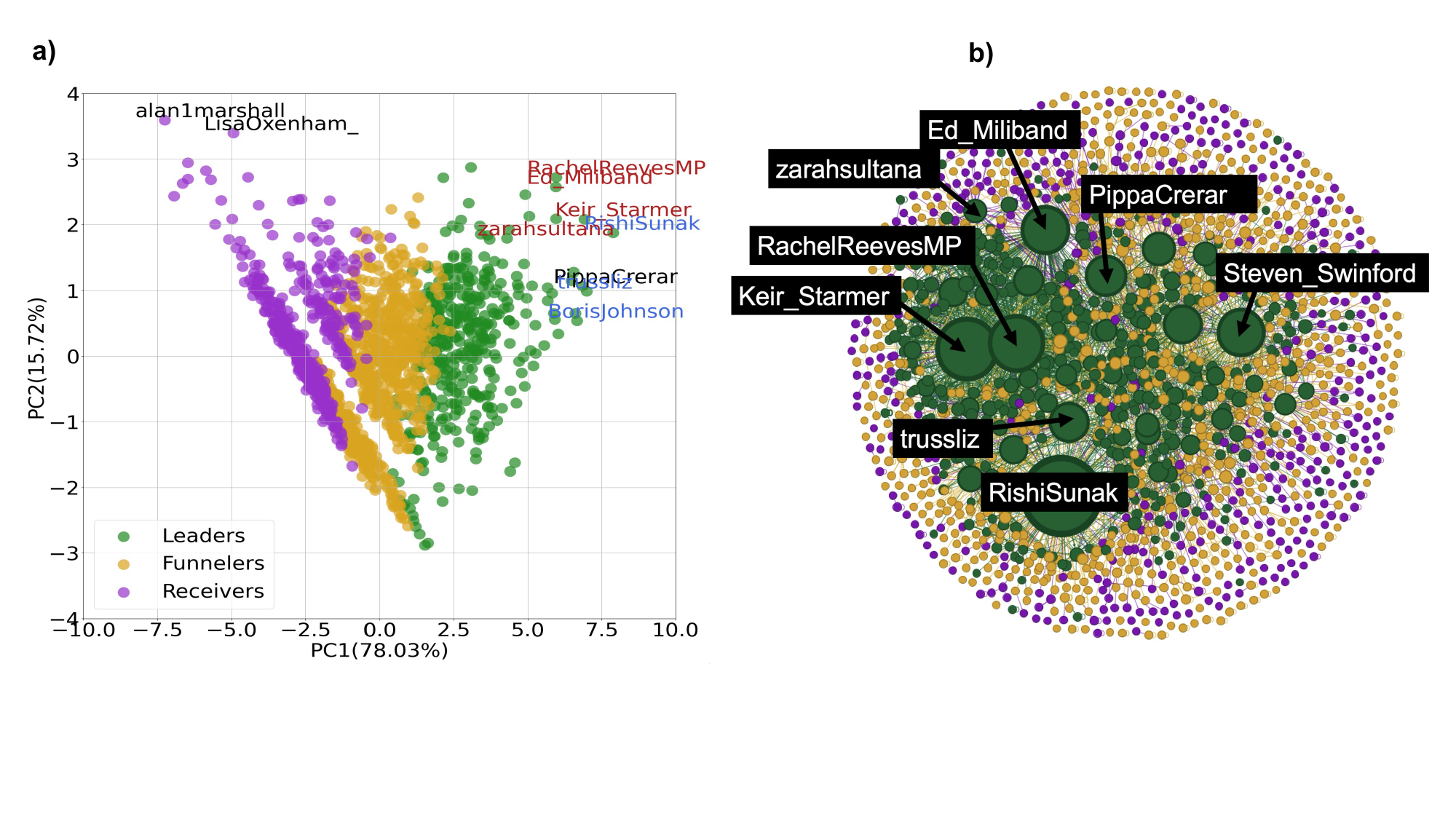}
     \caption{\label{fig:PCA-dimensions}PCA results for the cost-of-living crisis conversation. Figure (a) showing PCA results for the cost-of-living crisis topic. PC1 captures the inherent nodality of actors and PC2 captures the active nodalities. We cluster nodes using k-means clustering algorithm. We obtain 3 stable clusters that groups actors based on their inherent nodalities. Actors with high inherent nodality, are termed Leaders. Those with inherent nodality similar to that of the mean of the population are termed Funnelers and actors with less than average nodality are termed Receivers. Figure (b) shows the visualisation of the cost-of-living crisis network. We use the force-atlas algorithm for the layout. Nodes are coloured based on their cluster membership from the PCA. The size of the nodes reflects their weighted outdegrees. Actors with prominent parliamentary positions appear to be in the Leader cluster along with some other actors such as Zarah Sultana who has a back-bench parliamentary position but appear to have high active nodality on the topic (PCA results on the left).}
\end{figure*}

We visualise the cost-of-living-crisis network using these results. Each node is coloured based on the cluster they belong to (i.e., high, low, mean inherent nodality). We observe that the \emph{three} clusters form three separate tiers in the conversation network.
Nodes with high inherent nodality assume central positions in the conversation, i.e. we find that most of the interactions in the network are centred around these nodes. We term these nodes \emph{leaders}.
Nodes with less than average inherent nodality typically appear on the periphery of the network with limited connections to the rest of the network (average degree of 2). As these nodes appear to be at the periphery or at the receiving end of the information spread process, we term them \emph{receivers}.
Finally, we observe that nodes with average nodality (as the rest of the studied population) form the layer between the central nodes and the peripheral nodes. We find that these nodes appear to bridge the interactions between the \emph{leaders} and the \emph{receivers}, by receiving information from the \emph{leaders} (through retweets/mentions and replies) and passing it onto the \emph{receivers}, and hence we term them \emph{funnelers}.

\subsection{How does nodality explain influence on the platform?}
\label{ssec:group_nodality}
To examine how the relative nodality of an actor relates to their influence on the platform we employ a transfer entropy (TE) analysis which is independent of the conversation networks and studies the time-series of the actors' activities. More specifically, the TE analysis helps us measure the extent to which the activity of a group of actors can predict the activity of the rest of the actors in the conversation. In this instance, we study the groups of actors clustered based on their inherent nodality scores (as above) and perform a TE analysis across all four topics to examine how actors with different inherent nodalities influence the conversations. 

TE analysis typically captures the flow of information between two time-series and can be interpreted as the extent to which Y's current activity can be predicted by X's past. We determine the net information flow from $X$ to $Y$ as follows:
\begin{equation}
    \varphi(X, Y, t, k=1) = \frac{T_{X\to Y}(t, k=1)}{H_Y(t)}-\frac{T_{Y\to X}(t, k=1)}{H_X(t)},
\end{equation}
where $H_X(t)$ is the entropy of $X$ at time $t$. and similarly $H_Y(t)$ is the entropy of $Y$ at time $t$. The share of influence $\varphi$ at time $t$ gives the net flow of information between $X$ and $Y$ with a time-lag $k$, at time $t$. $\varphi$ lies on the interval [-1, 1]. If this quantity is positive, then there is an information flow from $X \rightarrow Y$, whereas if it is negative, information flows from $Y \rightarrow X$ \cite{schreiber_measuring_2000}.

\begin{figure*}[ht]
     \centering
     \includegraphics[width=\textwidth]{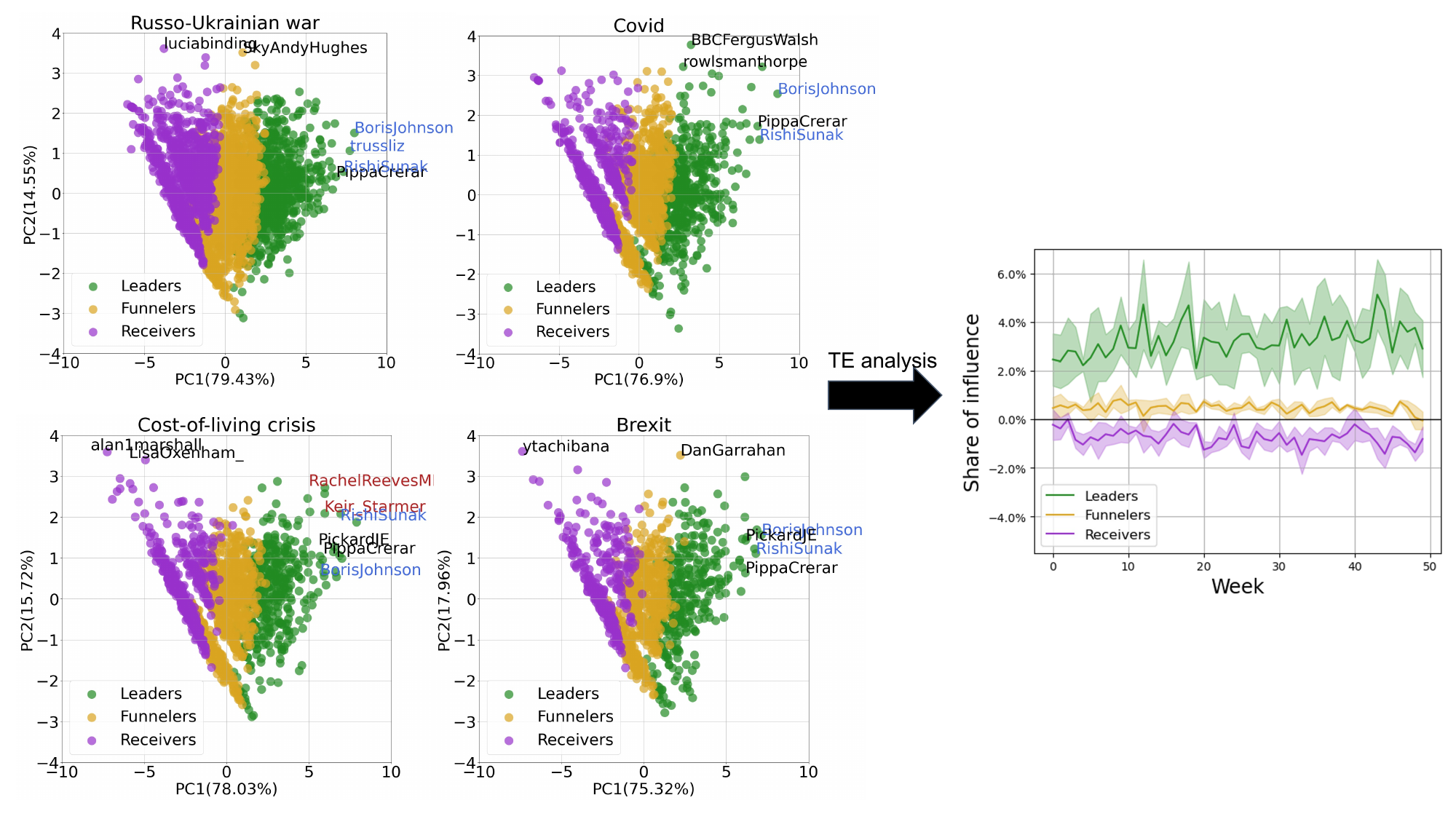}     \caption{\label{fig:Clusters}PCA results for all four topics. Figure on the left shows the PCA results for each of the four topics: the Russia-Ukraine War, Covid, cost-of-living crisis and Brexit. In each case, we run the k-means clustering algorithm where $k=3$. We find that this consistently groups actors based on their inherent nodality (i.e. along PC1). We then take the clusters of nodes for each topic and independently perform a transfer entropy (TE) analysis where we determine the impact of each cluster on the rest of the population. The results are then aggregated across all four topics and presented in the figure on the right. We find that the Leader cluster on average has higher influence than the other two clusters on the activity of the population across all four topics.}
\end{figure*}

The results of the TE analysis aggregated across all four topics are shown in Figure \ref{fig:Clusters}. 
We find that \emph{leaders}, i.e. the group of actors with high inherent nodality consistently influence the rest of the population positively.
\emph{Receivers}, on the other hand, have a negative share of influence, which indicates that they mainly respond to the ongoing activity in the network and primarily receive information. Whereas \emph{funnelers} demonstrate low positive values suggesting that while their activity may influence the activity of the \emph{receivers}, their own activity is influenced by the activity of the \emph{leaders}. This suggests that nodality can be a useful tool to study the roles of actors in online discourse, and help determine who has the capacity to shape the discourse (by influencing the activity related to the topic).

We now use the approach proposed above to answer our original research question \textemdash who between the MPs and journalists leads the conversation on the policy topics studied here? Specifically, we want to determine who is leading whom (i.e. MPs leading journalists or vice-versa) in each of the four discussions.
For a more effective analysis, we split MPs and journalists into sub-groups to reflect their organisational roles (in order to capture influence related to institutional nodality). 
We exploit the structure of the British parliament and classify MPs into four groups: the Cabinet, which forms the executive power; the Shadow Cabinet, formed by the Opposition party that oversees the activities of the Cabinet; the Backbench, formed by MPs of the ruling party who do not have a place in the Cabinet; and its analogous for the opposition party, simply called the Opposition.
We also divide journalists into two groups based on their follower counts. We find that the ratio of the Cabinet to the Backbench MPs and of the Shadow Cabinet to the Opposition (backbench) MPs is roughly 10\%. We also split the journalists by separating the 10\% most-followed journalists from the rest of the journalists (and term them prominent journalists).

\begin{figure*}[ht]
     \centering
     \includegraphics[width=\textwidth]{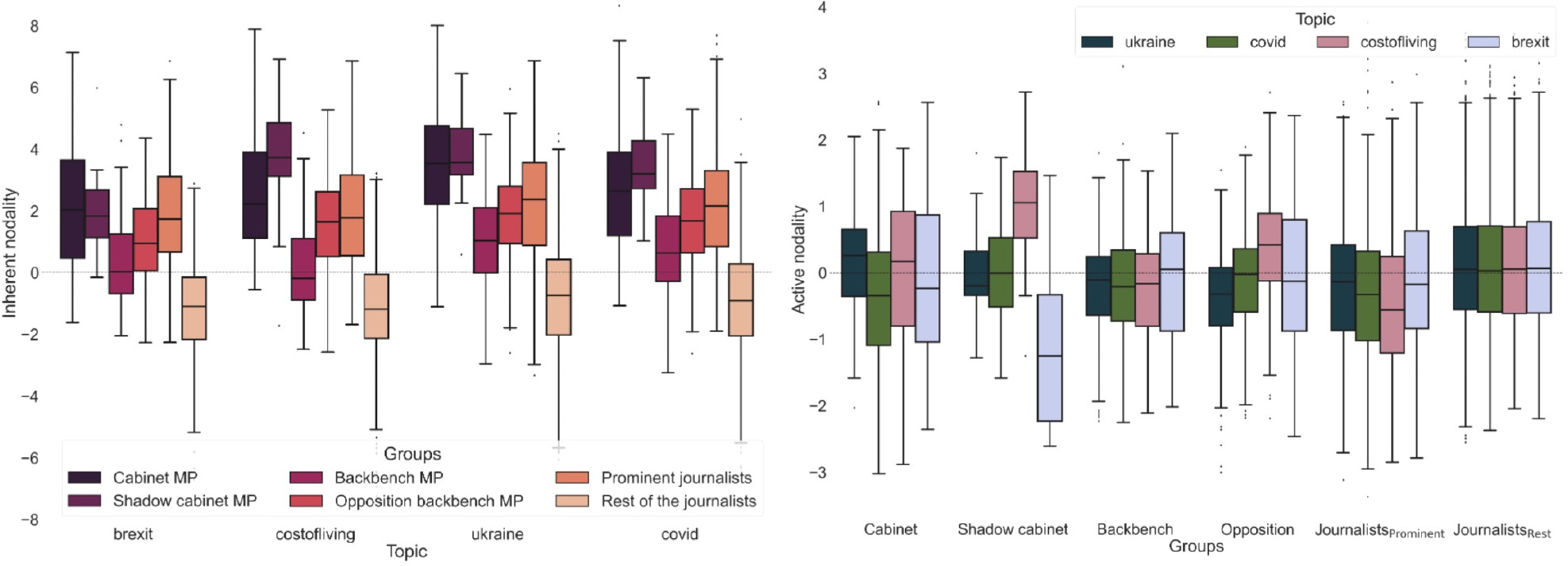}
     \caption{\label{fig:boxplots}Inherent and active nodality of all sub-groups of actors across all four topics. The boxplot on the left shows the inherent nodality of actors for each sub-group and on each topic. We find similar trends across all topics, where MPs with key parliamentary positions have highest inherent nodalities, followed by the most followed journalists and then the rest of the actors. The boxplot on the right summarises the active nodality of the sub-groups. We obtain more nuanced results where different sub-groups of actors have different rankings across topics.}
\end{figure*}

Next, we repeat the PCA analysis as above, but assign the nodality scores as the mean values for each group. 
Results of the PCA are shown in Figure \ref{fig:boxplots}. 
We find that Cabinet MPs and Shadow Cabinet MPs have the highest inherent nodality for all four topics, along with prominent journalists. This implies that actors with higher institutional status inherently have more visibility and engagement from the other actors on each of the four topics.
On examining the active nodality of actors, we find that on average more Shadow Cabinet ministers have nodality specific to the topic of cost-of-living crisis and limited active nodality on the topic of Brexit. 
We further observe that the Opposition back-bench has relatively high active nodality on the cost-of-living crisis topic, suggesting that the Opposition Party as a whole had more engagement from journalists and MPs on the platform on the issue of the economic crisis under the Conservative government, which also went onto being a focus of the Labour Party's election manifesto in 2024 \cite{stafford2024}.

So far both inherent and active nodalities have been found to play an important role in the level of engagement an actor receives on a given topic on the platform. However, does one of the dimensions matter more than the other in driving influence on the platform?

To analyse the association between the share of influence and the two types of nodality, we perform a regression analysis given by:

\begin{align}
    \varphi(G, \neg G, t) = \ \mathrm{constant} + \gamma t + \sum_{i\in G}&\alpha_\mathrm{inherent} \mathrm{PC}1_i^t+ \sum_{i\in G}\alpha_\mathrm{active} \mathrm{PC}2_i^t \nonumber \\
    &+\sum_{i\in G} \beta \mathrm{PC}1_i^t\times \mathrm{PC}2_i^t, 
    \label{eq:linear_model}
\end{align}
where $G$ refers to a given group, which in our case is the Cabinet, the Shadow Cabinet, etc. $\neg G$ refers to the complement of $G$, thus all nodes not belonging to $G$. PC1 and PC2 refer to the first two components of the resulting PCA. $\alpha_\mathrm{inherent}$, $\alpha_\mathrm{active}$ and $\beta$ are three independent coefficients. The combined term PC1$\times$PC2 is included to determine any interaction effect. A temporal term $\gamma t$ is also included to capture the effect of time.
We fit the LHS of Eq. (\ref{eq:linear_model}) with the influence score of each group from the TE analysis. The RHS of the equation is fitted with the mean inherent and active nodalities at group level, obtained from the PCA. We standardise the time-period across the TE analysis and the PCA, by creating snapshots of networks over a period two weeks (for the PCA), which is also the time-lag considered in the TE analysis. 




\begin{table}[h]
\centering
\caption{\label{tab:results_linear_model}Results of the linear regression.}
\begin{tabular}{l|c|c}
    \toprule
    \multicolumn{3}{c}{$N=1045, \ R^2=0.6137$} \\
    \midrule
    Variable & Coefficient & $p$-value \\
    \midrule
    Inherent*** & $0.010 \pm 0.001$ & 0.000 \\
    Active* & $0.008 \pm 0.004$ & 0.025 \\
    \makecell[tl]{Inherent-Active} & $-0.001 \pm 0.001$ & 0.111 \\
    Time*** & $-0.002 \pm 0.000$ & 0.000 \\
    Intercept** & $0.088 \pm 0.029$ & 0.002 \\
    \bottomrule
\end{tabular}
\end{table}

Results of the regression analysis are presented in Table \ref{tab:results_linear_model}. We make three key observations: 
(i) the capacity of a group to influence the discussion related to a topic depends both on their engagement on the topic as well as their institutional status; (ii) the capacity of a group to influence topic-related activity is positively related to the mean inherent and active nodalities of the group; and finally, (iii) we also show that the \emph{two} dimensions of nodality (inherent and active) are independent of each other, since the crossed term (inherent-active) does not contribute to the capacity of the group to influence activity. Thus establishing a robust methodology to quantify nodality of actors along two independent dimensions \textemdash (i) institutional status related nodality (or inherent nodality) which is transferable across topics and (ii) topic-specific nodality (or active nodality).
\section{Discussion}
\label{sec:discussion}

To date, the concept of nodality has been studied largely from a theoretical perspective \cite{margetts_nodality_2023, castelnovo_nodality_2021, howlett_government_2009, howlett_information_2022}, with very few attempts to quantify nodality in an empirical setting \cite{margetts_governing_2006, petricek_web_2006}. Here for the first time, we focus on computing nodality scores for individual actors in policy conversations as a tool to gauge their relative influences in policy-related online discourse.
More specifically, for our two sets of elite actors (MPs and journalists), we distinguish and measure two dimensions of nodality; (i) topic-specific active nodality at the individual level, relating to the activity of an actor on a specific topic, and (ii) inherent nodality relating to the actor's institutional role and status. We show that distinguishing between the two types of nodalities provides a useful way to determine the \emph{nodal advantage} of actors in prominent roles, such as in the case of Ministers and political editors, who by virtue of their institutional status possess nodality which can be transferred across topics. Our main finding is that, while both institutional status and platform activity is important, activity can generate limited topic-specific influence whereas institutional status tends to bestow influence across topics.

This work enriches our understanding of the role of policy-actors in policy-related conversations, as it goes beyond studies which look at the influence of individuals in a conversation only in terms of their network attributes or positions in the network (i.e. centrality metrics) \cite{dubois_multiple_2014, riquelme_measuring_2016, esteve_del_valle_leaders_2018, zhang_exploratory_2023}. Indeed, we introduce a methodology that measures the \emph{capacity} of an actor to influence others in an online policy conversation \textemdash either on the basis of their institutional status (which is platform- and topic-independent), or as an emerging leader with growing topic-specific nodality.
Our study also expands on the literature of politics and journalists on Twitter \cite{mosleh_measuring_2022, parmelee_political_2013, romenskyy2018polarized, jungherr_twitter_2014, kim_measuring_2012, graham_between_2013, graham_role_2015, bovet_influence_2019, nozewski_birds_2021, shapiro_politicians_2017, bright_explaining_2018, mosleh_misinformation_2024, li2024quantifying}, introducing the actor’s role in their institutions and their individual activity on a topic as key elements in understanding the influence they exert. 

Going forward, our methodology could support further research investigating the activities and influence of politicians on online platforms. For instance, here we classify \emph{inherent} nodality as pertaining to institutional status, but there are other potential drivers of inherent nodality that our approach might help to identify, such as educational background, that could contribute to their `eliteness' by virtue of being within specific social circles that in turn could then drive their nodality on the platform, independent of individual levels of activity.  Alternatively, our methodology might assist the study of different, more negative kinds of attention, such as hate speech directed at political representatives, where our approach might help assess which element of the hate directed at ministers relates to their institutional role or their individual characteristics. This has been a challenge to examine in the past \cite{gorrell_which_2020}, given that ministers are more well-known and hence get more hate, regardless of race or gender \cite{agarwal_hate_2021}. Likewise, for journalists, who are increasingly suffering threats of violence online, such an approach might help determine whether the hate they receive relates to the publications they work for, or to their historic activity on policy-topics such as gender.

Finally, there are some limitations to this work. First, our elite networks include only British journalists and MPs. This is a limited set of individual actors, albeit an important one in terms of nodes that have both individual and institutional nodality, and who play an important role in setting the policy agenda. Second, we have covered only four policy topics. Future work might focus on larger domains, such as defence and security (rather than the Ukraine war as focused on here), which could allow us to say something generalisable about elite networks in broader domains. Third, our results represent an analysis of the volume of activity – and not the content, which could also be insightful \cite{jimenez2021sentiments}. Fourth, we have looked at only one specific social media platform - Twitter, now X, a platform currently experiencing a decline\footnote{\href{https://variety.com/2023/digital/news/musk-twitter-x-acquisition-one-year-user-revenue-decline-1235770297/}{variety.com/2023/digital/news/musk-twitter-x-acquisition-one-year-user-revenue-decline-1235770297/}}, which may not be possible to analyse in this way in the future, due to the lack of data and also lack of universality (due to the 2024/25 exit from X). It will be increasingly important therefore, to conduct such research on different platforms.

%

\section*{Acknowledgments}
 The authors would like to thank Angus R. Williams for providing the first list of Twitter handles associated with MPs. The authors would also like to thank Eirini Koutsouroupa for her support and effort to get this work published.


\section*{Funding}
This work was supported by the Public Policy Programme of The Alan Turing Institute under the EPSRC grant EP/N510129/1.
This work was supported by the Ecosystem Leadership Award under the EPSRC Grant EP/X03870X/1 \& The Alan Turing Institute.

\section*{Author contributions statement}
L.CG., S.C., H.R., H.M., D.G. and J.B. conceived the experiment(s),  L.CG. and S.C. conducted the experiment(s), L.CG., S.C., H.R and H.M. analysed the results. L.CG., S.C. and H.M. wrote the manuscript and L.CG., S.C., H.R., H.M, D.G and J.B. reviewed the manuscript.


\section*{Data availability}
Twitter data is made available in accordance with Twitter’s terms of service. Tweet  IDs and analysis code are available at \url{https://github.com/Turing-Online-Safety-Codebase/nodality}. The corresponding tweets can be downloaded using the official Twitter API (\url{https://developer.twitter.com/en/docs/twitter-api}).

\section*{Appendix}
\label{sec:appendix}

\subsection{Weak supervision classifier}
\label{ssec:appendix-weak_supervision}
To classify tweets by topic, we build a classifier using a weak supervision approach. This allows us to move beyond keyword search and create heuristic rules tailored to specific topics. Weak supervision has found application in various contexts such as medicine and health sciences \cite{Fries2021, Silva_2022}, chemistry \cite{Mallory_2020}, detection of fake news \cite{Shu_2021}, and more generally to study sentiment analysis \cite{Jain_2021, castro-gonzalez_cheap_2024}, and ontology and computational linguistics problems \cite{Maresca_2021, Berger_2021}.

The classifier is based on a set of labelling functions (LFs) and a probabilistic model that applies these functions to the unlabelled data. 
LFs are rule-based functions that assign labels based on specific criteria, such as keywords, regular expressions, Named-Entity Recognition (NER) systems, and NLP features like polarity and subjectivity. This diverse set of criteria results in topic-specific LFs that are tailored to the nuances of each topic, offering a more sophisticated approach than simple keyword searches. 
Given $m$ labelling functions and $n$ unlabelled data points, the process of applying all labelling functions to the entire data set will result in an $n\times m$ matrix $\boldsymbol{M}$ where each entry reflects how a given labelling function acted on a data point. 
By having a training data set with golden labels $\boldsymbol{Y}$, the probabilistic \href{https://snorkel.readthedocs.io/en/v0.9.3/packages/_autosummary/labeling/snorkel.labeling.LabelModel.html}{\texttt{LabelModel}} \cite{Ratner2019} computes the conditional probability of labelling functions given the matrix $\boldsymbol{M}$, $P(LF|\boldsymbol{M}, \boldsymbol{Y})$. The latter probabilities are used as parameters to then obtain a final label for each data point. 

In our case, we build a collection of labelling functions for each one of the four studied topics. A detailed list of these functions can be found in our GitHub repository\footnote{\url{https://github.com/Turing-Online-Safety-Codebase/nodality}}. With the purpose of illustrating how labelling functions are created, we give two examples of pseudo LFs for the topic of the Cost of Living Crisis:
\begin{align*}
    \mathrm{LF1:} &\ \texttt{label CRISIS if any keyword in keywords in datapoint else ABSTAIN}. \\
    \mathrm{LF2:} &\ \texttt{label CRISIS if `(rising|inflation|rise)\textbackslash W+(?:\textbackslash w+\textbackslash W+){0,6}?} \\ &\texttt{(inflation|is rising)' in datapoint else ABSTAIN} \\
\end{align*}
While LF1 refers to a normal keyword search, LF2 refers to a more complex regular expression regarding the rise of inflation. LF2 allows us to look for different forms of the same sentence ``The inflation is rising''.

\subsubsection*{Confusion matrix of the classifier}

\begin{table}[ht]
    \centering
    \caption{\label{tab:confusion_matrix}Confusion matrix of the classifier. The size of the test data sample is $N=598$.}
    \begin{tabular}{c|c|c|c|c|c|c}
    \toprule
    \multicolumn{7}{c}{Actual Topic} \\ 
    \cmidrule{2-7}
    \parbox[t]{2mm}{\multirow{6}{*}{\rotatebox[origin=c]{90}{Predicted Topic}}} & & Other &  Ukraine & COVID-19 & Cost of Living Crisis & Brexit \\
       \cmidrule{2-7}
        & Other & 8.99e-1 & 1.65e-2 & 2.27e-2 & 5.98e-2 & 2.00e-3 \\
       & Ukraine & 0.00 & 1.00 & 0.00 & 0.00 & 0.00 \\
       & COVID-19 & 0.00 & 0.00 & 1.00 & 0.00 & 0.00 \\
       & Cost of Living Crisis & 0.00 & 0.04 & 0.00 & 0.96 & 0.00 \\
       & Brexit & 0.00 & 0.00 & 0.25 & 0.00 & 0.75 \\
       \bottomrule
    \end{tabular}

\end{table}



\bibliographystyle{apalike}
\bibliography{main}
\end{document}